# Robust predictive model for Carriers, Infections and Recoveries (CIR): first update for CoVid-19 in Spain

Efrén M. Benavides

Universidad Politécnica de Madrid

April 11, 2020

## I. Introduction

With the purpose of predicting the evolution of infectious diseases under uncertainty or low-quality information, just as happened in the initial scenario during the spread of CoVid-19 in China and Europe, a new model has been presented in [1]. To achieve this goal, the model implements the following four key characteristics:

1. It keeps track of the date of infection of a single individual.
2. It uses stochastic distributions to aggregate individuals who share the same date of infection.
3. It uses two types of infections: mild and serious.
4. It keeps track of the number of **C**arriers, **I**nfections and **R**ecoveries.

The kernel of the presented model consists of four differential equations: two for Infections and two for Recoveries. This kernel is complemented with additional differential equations, one for each sanitary event to be predicted: among others, deaths and demands of ICUs or mechanical ventilators under different restraint policies.

Since there are new data available, this paper revises the model and the prediction made in [1] using the new data reported by reference [2] and incorporating also data from reference [3].

The calculations in this work are made with the only purpose of assessing the model: input parameters may require further study and predicted results might suffer changes. However, it is anticipated that the proposed predictive model holds good agreement with reported data.

## II. Differential equations

The kernel of the proposed model, detailed in reference [1], is condensed as follows

$$\frac{dI_i(t)}{dt} = \bar{\omega}\gamma\phi_i \left[1 - \sum_i \frac{I_i(t)}{P}\right] \sum_i (1-\alpha_i) C_i(t) \qquad (1)$$

$$\frac{dR_i(t)}{dt} = \int_0^t \frac{dI_i(t-\Delta t)}{dt} h(\Delta t, \mu_{Ri}) d\Delta t \qquad (2)$$

In these equations, subindex $i$ indicates the type of infection (in this case, 1=mild, 2=serious); $C$ is the number of carrirers, $I$ is the number of infections and $R$ is the number of recoveries; finally, $P$, $\alpha_i$, $\bar{\omega}$, $\gamma$ and $\phi_i$ are model constants which can be selected to model different scenarios or restriction policies.

- $P$ is the susceptible population. It takes into account that there are people who are not accessible to the contagion. The susceptible population changes when regions

- are added to the contagion path: this fact can be considered by increasing the value of $P$ at any time.
- $\alpha_i$ is the proportion of carriers of type $i$ who are isolated (at home or hospital) and cannot propagate the virus: the carriers propagating the virus are $\sum_i (1-\alpha_i)(t)$ with $0 \leq \alpha_i \leq 1$. When there are two types, it is possible to define $\omega = (1-\alpha_1)\bar{\omega}$ and $\alpha = (\alpha_2 - \alpha_1)/(1-\alpha_1)$, which are the parameters proposed in [1]. It is convenient to clear $\alpha_1 = (\alpha_2 - \alpha)/(1-\alpha)$ and $\alpha_2 = \alpha_1 + \alpha(1-\alpha_1)$ to find that $-\infty < -\alpha_1(1-\alpha_1)^{-1} \leq \alpha \leq \alpha_2 \leq 1$. When $\alpha \leq 0$, $\alpha_2 \leq \alpha_1$ holds, that is, type-1 (mild) infections are more isolated than type-2 (serious) infections and/or that serious infections are more contagious than mild ones. As long as these parameters are obtained by means of a fitting algorithm, only two parameters are needed: $\omega$ and $\alpha$ (instead of $\bar{\omega}$, $\alpha_1$ and $\alpha_2$).
- $\bar{\omega}$ is the number of persons that an average non-isolated carrier finds per day. As said, for fitting purposes, it is better to use $\omega = (1-\alpha_1)\bar{\omega}$, which is the number of persons that an average carrier finds per day. It takes into account how people move and come closer.
- $\gamma$ is the contagion success. It measures how many persons in a healthy group become infected when keeping in physical proximity with an infected person. It takes into account the efficiency of the transmission channel: the use of masks, for example, will reduce this number.
- $\phi_i$ is the fraction of infections generated of type $i$. In a group of new infected people, it measures how many of them will be of each type. For two types, $\phi_2 = 1 - \phi_1 = \phi_r$ is the fraction at risk.
- $h(\Delta t, \mu_{Ri})$ is the stochastic distribution for each event and it is described in [1], where preliminary values were assumed as a first attempt. So far, it has not been necessary to change or improve it.

The model is completed with the differential equations associated to the medical events (first symptoms, ICU admission, discharging, death, etc.). In general, these equations are of the following form

$$\frac{dIH(t)}{dt} = \int_0^t h(\Delta t, \mu_{IH}) dX_j(t_0) = \int_0^t \frac{dX_j(t-\Delta t)}{dt} h(\Delta t, \mu_{IH}) d\Delta t \qquad (6)$$

$$\frac{dOH(t)}{dt} = \int_0^t h(\Delta t, \mu_{OH}) dIH(t_0) = \int_0^t \frac{dIH(t-\Delta t)}{dt} h(\Delta t, \mu_{OH}) d\Delta t \qquad (7)$$

$$H(t) = IH(t) - OH(t) \qquad (8)$$

In these equations, $IH(t)$ is the number of people which is an input for the service $H$, $OH(t)$ is the number which is an output for the service $H$ and $H(t)$ is the people who are using the service $H$. The function $X_j$ denotes one of the previous calculated functions: $X \in \{C, I, R\}$ and $j \in \{1,2\}$. In reference [1] it was assumed that only type-2 carriers ($X = I$ and $j = 2$) become users of the medical service and the same is assumed here, $X = I$ and $j = 2$. In addition, following [1], deaths are a constant fraction $t_D$ of the people leaving the hospital, so that

$$D(t) = t_D OH(t) \qquad (9)$$

## III. Previous results

Reference [1] studied three different initial dates and shown that, as expected, the longer the time available for viral spreading, the greater the exposed population. Moving the date from February 5th to January 25th (eleven days) increases the susceptible population from 0.129 million to 13.8 million: a change of two orders of magnitude! This fact gives a tremendous importance to the date of the first infection because it fixes the number of immune people, which is crucial to determine the contagion spreading after removing the mobility restrictions. Without knowing the number of people who have passed the infection, that is, without testing the population to determine the fraction who has developed immunity, this number cannot be definitely established, it can only be guessed. The work in reference [1] stated the plausible initial date as February 3rd (please, note that this is a plausible date, not a fact). This plausible scenario is drawn again in Figure 1, where the new available data has been added. The figure shows that the trend is good although the real data is rising above the estimation. Next sections discuss this situation.

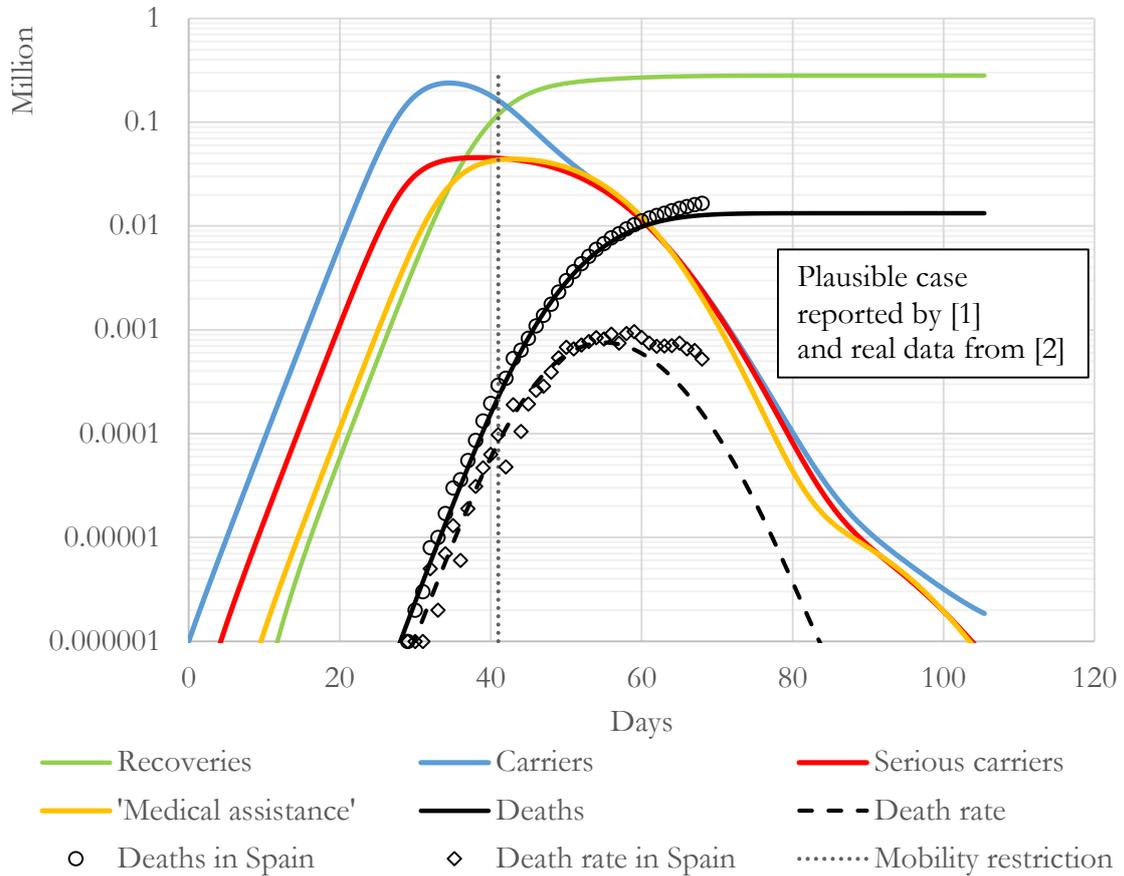

FIGURE 1. Spreading of CoVid-19 in Spain assuming February 3rd as the initial day. Solid lines are calculated using the CIR model with the input parameters: $\mu_{IH}$ = 3.10 days, $\mu_{OH}$ = 11.36 days, $\mu_{I1}$ = 6.72 days, $\mu_{I2}$ = 13.92 days, $t_D$ = 0.283, $\gamma$ = 0.165, $P$ = 0.2833 million, $\alpha$ = 0.550, $\omega$ = 2.95 and $\phi_2$ = 0.167. Real data comes from reference [2].

## IV. Comparison with new real data

Real data in the previous work was obtained from reference [2], which has the advantage of reporting the death rate every day since the beginning. Although they use official data they are not official; the official data comes from the Spanish government [3]. However, reference [3] does not give the death rate, that is, the number of deaths per day, but reports, without stating when the death happened, the accumulated deaths until such date. Thus, as reference [3] advises: death rates cannot be directly extracted from the given data and anybody who will use it for such purpose will do so at their own risk. In addition, the first 28 deaths are all accumulated on March 8$^{th}$, which reduces the information available for estimating the trending slope during the first days. Figure 2 shows the difference between both sets of data (reference [3] minus reference [1]).

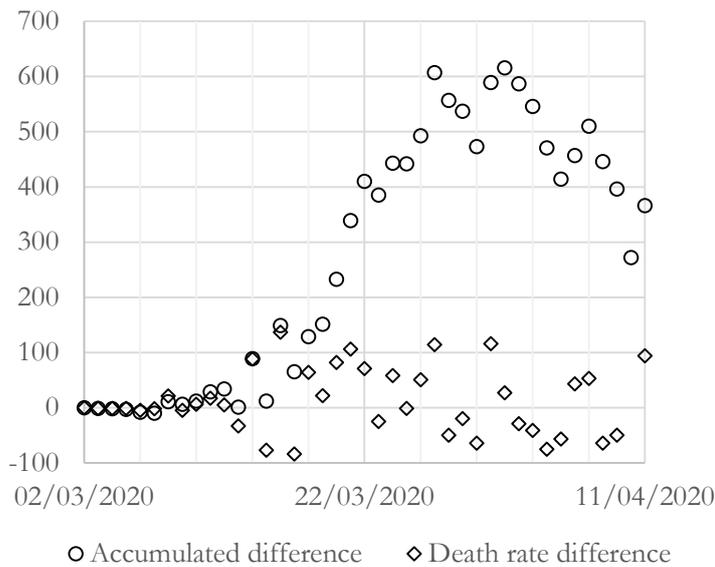

FIGURE 2. Differences between death rates and deaths reported by references [3] and [2]. Both curves represent data from [3] minus data from [2].

As can be seen, on April 11$^{th}$ there exists a difference of 366 deaths but the difference was as high as 616 deaths on April 1$^{st}$: reference [2] has a delay reporting the deaths. This partially explains why the real data is going above the calculated curve in figure 1: as long as reference [2] reports less deaths, the parameters fitted by [1] slightly underestimate the data reported by [3]. This can be seen in figure 3.

Figure 3 shows the real data from reference [3] where the lack of data during the first days appears. This lack of data contributes to make less accurate the determination of the plausible initial date. However, the rising of deaths at the end is significantly high. As said, a partial explanation comes from the difference between the deaths reported by references [3] and [2] but there must exist another reason. Three plausible reasons could be argued:

1. The proposed mathematical model does not properly describe the propagation.
2. There exists a significant delay reporting the deaths: there are deaths reported on one date which happened on a previous one.
3. There exists an increment of the available population: for some reason, areas of population which were hidden to the contagion are suddenly connected to the infectious path.

Option 1 is discarded since the model has been written to avoid this possibility: an effort was made to incorporate all the major effects; only minor effects could be added, and they would not change significantly the results. Option 2 cannot be checked and hence it is also discarded. Option 3 could be plausible since on March 8[th] there were many demonstrations that could have spread the virus to isolated regions. Next section re-adjusts the model parameters to incorporate this possibility.

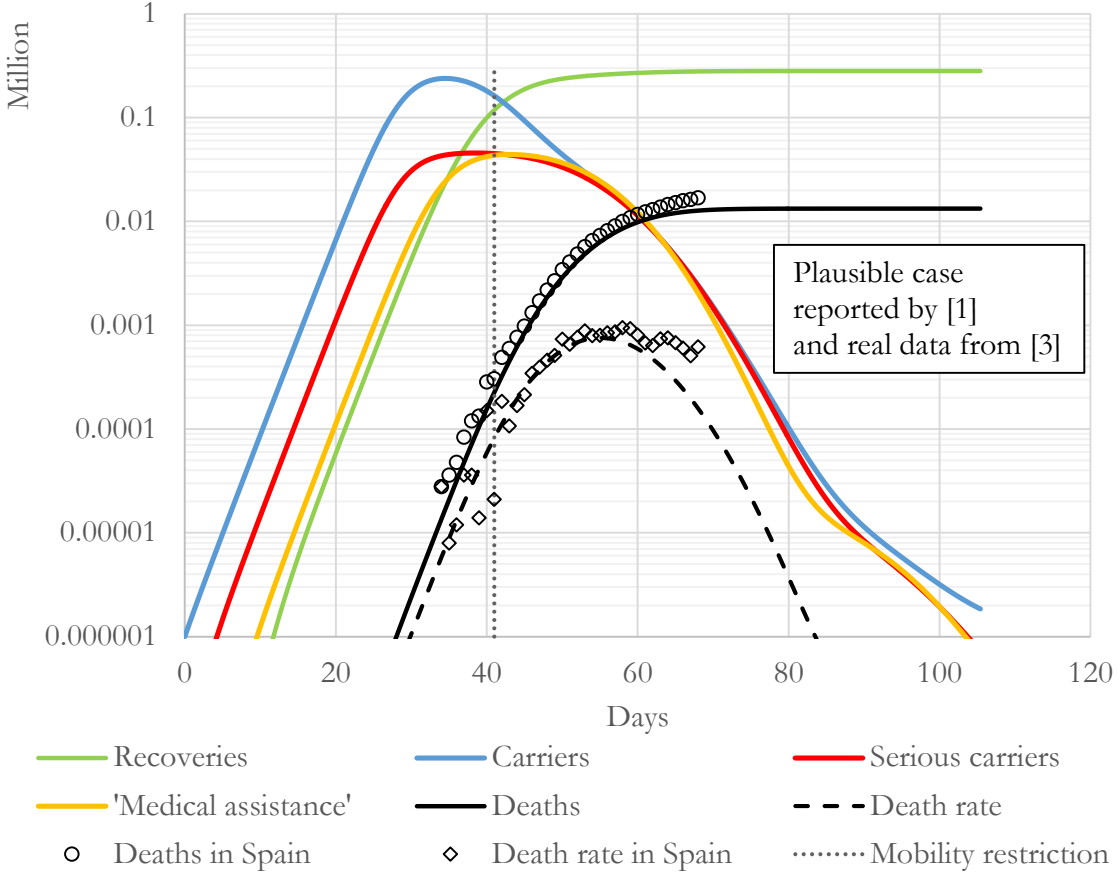

FIGURE 3. Spreading of CoVid-19 in Spain assuming February 3[rd] as the initial day. Solid lines are calculated using the CIR model with the input parameters: $\mu_{IH}$ = 3.10 days, $\mu_{OH}$ = 11.36 days, $\mu_{I1}$ = 6.72 days, $\mu_{I2}$ = 13.92 days, $t_D$ = 0.283, $\gamma$ = 0.165, $P$ = 0.2833 million, $\alpha$ = 0.550, $\omega$ = 2.95 and $\phi_2$ = 0.167. Real data comes from reference [3].

## V.  New fitting

The data from reference [3] has been used to re-estimate the model parameters. This fitting has been made for three plausible initial dates, February 3[rd], February 5[th] and February 7[th].

The restriction of movements imposed by the government is described with a reduction of $\omega$ in such a way that after March 15[th], one day after its official publication, $\omega$ is divided by $r_1$ (reference [1] used $r_1$ = 10). On April 3[rd] the restriction was tightened (following the new regulations imposed by the government on April 2[nd]) dividing $\omega$ by $r_2 > r_1$.

The effect of increasing the available population which is susceptible to contagion is incorporated using two values of $P$: $P_1$ before March 9[th] and $P_2$ after March 8[th].

The values of the parameters $P_1$, $P_2$, $\alpha$, $\omega$, $r_1$, $r_2$ and $\phi_2$ which minimize the logarithmic error are collected in Table 1. The rest of the parameters remain equal to those estimated by reference [1].

As long as the number of available empirical data is greater, the logarithmic error is also larger, near 4.2 for the three cases (the error in [1] was near 2.1). The differences in the error obtained in the three cases are less than 0.5%, what makes very difficult to use this error to determine the initial date.

| First infection | February 3$^{rd}$ | February 5$^{th}$ | February 7$^{th}$ |
|---|---|---|---|
| $P_1$ (million) | 21.13 | 5.541 | 1.406 |
| $P_2$ (million) | 48.82 | 12.68 | 3.111 |
| $\alpha$ | 0.173 | 0.174 | 0.181 |
| $\omega$ (pers./pers./day) | 4.060 | 4.056 | 4.074 |
| $r_1$ | 7.752 | 5.728 | 4.492 |
| $r_2$ | 10 | 10 | 10 |
| $\phi_2$ | 0.0016 | 0.0063 | 0.0236 |
| Log. Error (total) | 4.205 | 4.205 | 4.185 |
| $R(\infty)$ (million) | 46.64 | 12.20 | 3.016 |
| $\phi_D(\infty)P$ | 21766 | 21788 | 20118 |

TABLE 1. Set of parameters found by minimizing the total logarithmic error using the data given by reference [3]. The last two rows give the stationary result for recoveries and deaths.

As can be seen, the change of ±2 days in the initial date produces a change in the susceptible population of several millions, from 3.11 (February 7$^{th}$) to 48.8 (February 3$^{rd}$). As discussed in [1], the parameter $\phi_2$ suffers similar changes, from 0.236 to 0.0016. This is a severe change respect to the value 0.167 estimated in reference [1]. The rest of parameters have not significantly changed with respect to those of reference [1]: $\omega$ has increased and $r_1$ decreased.

In the three cases, the fitting algorithm has increased the susceptible population from 1.4 million to 3.1 million using February 7$^{th}$ as the initial date and from 21.1 million to 48.8 million using February 3$^{rd}$ as the initial date. Again, without testing the population it is difficult to infer the people who have developed immunity and hence the initial date has to be indirectly determined. In this case, as long as February 7$^{th}$ leads to a value of $\phi_2$ closer to the plausible limits which reference [1] estimated, it seems a more plausible initial date than February 3$^{rd}$. This fixes, as a plausible option, 3 million people who have passed the infection and 20 thousand final deaths.

Next figures show the calculated curves for the three initial dates. It is interesting to compare them with the calculation on figure 3. One of the changes is the lower slope that the death rate curve has after its maximum. The other effect is the higher duration of the decay, which is near one week longer.

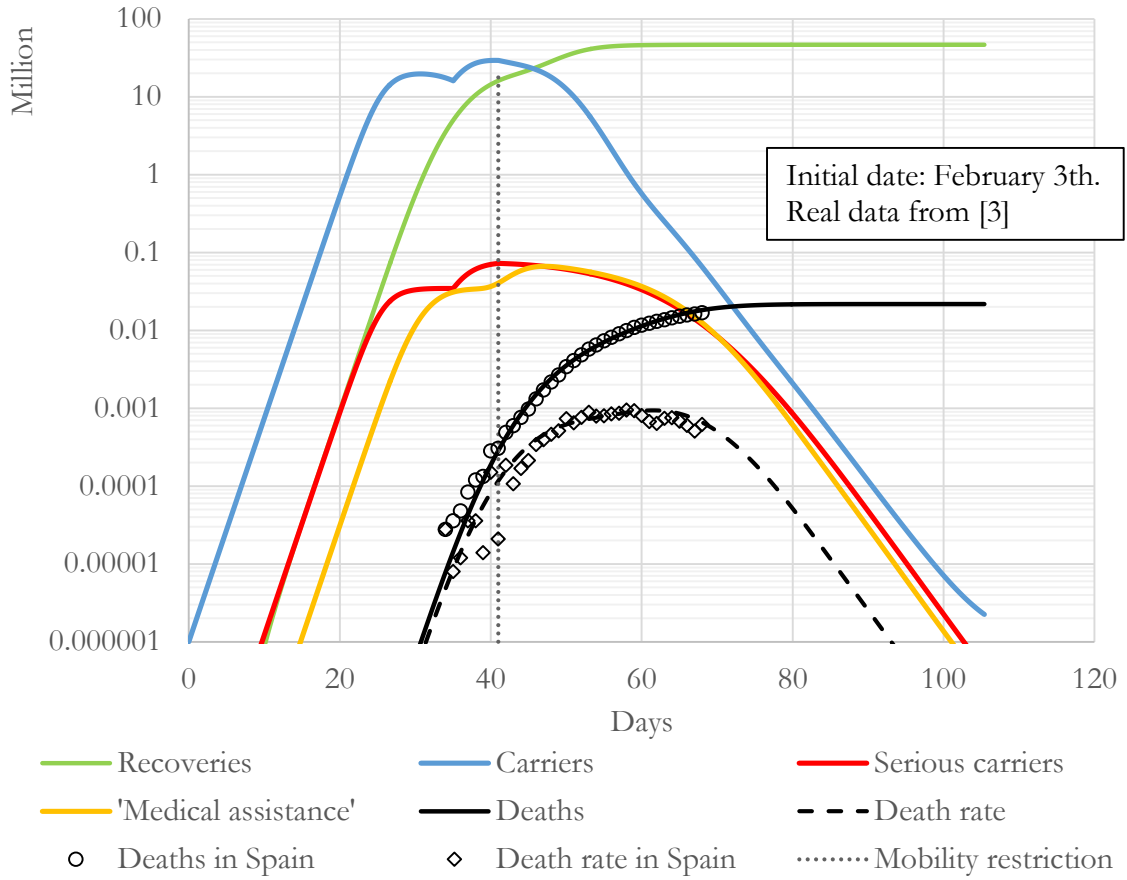

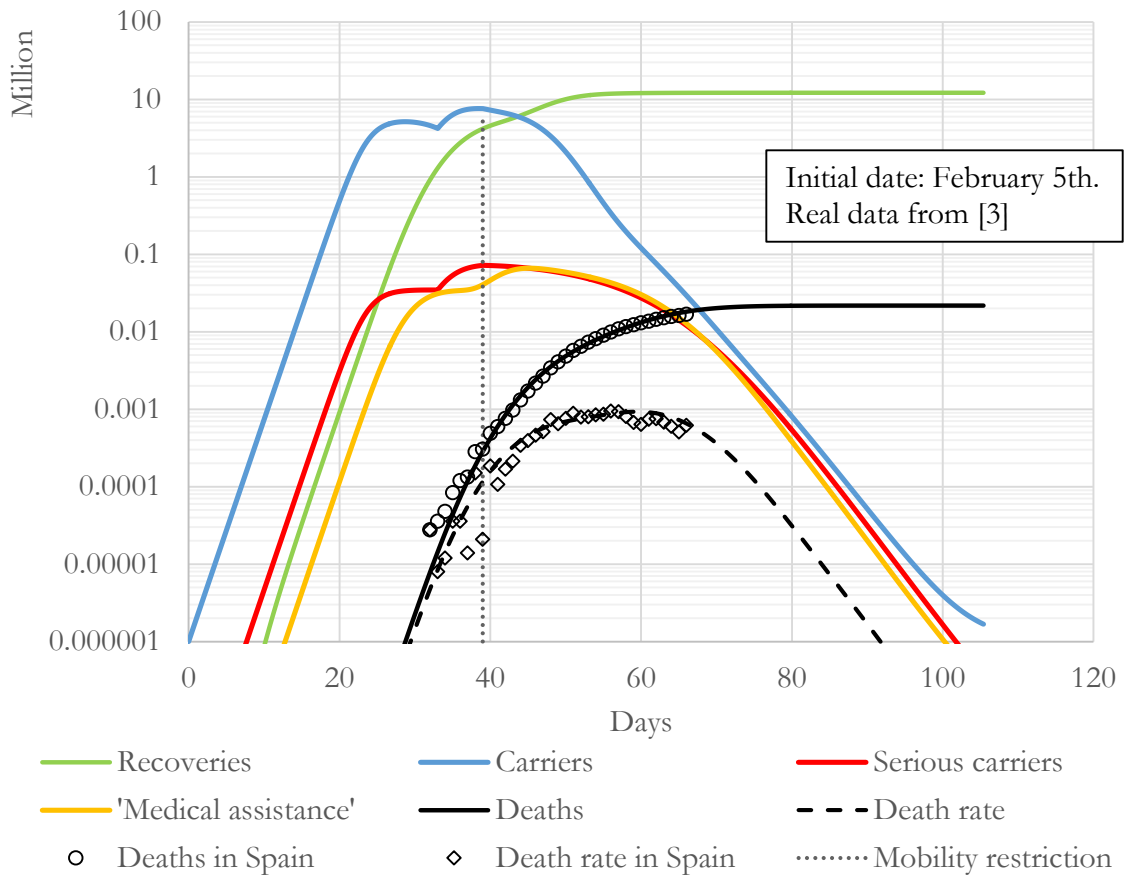

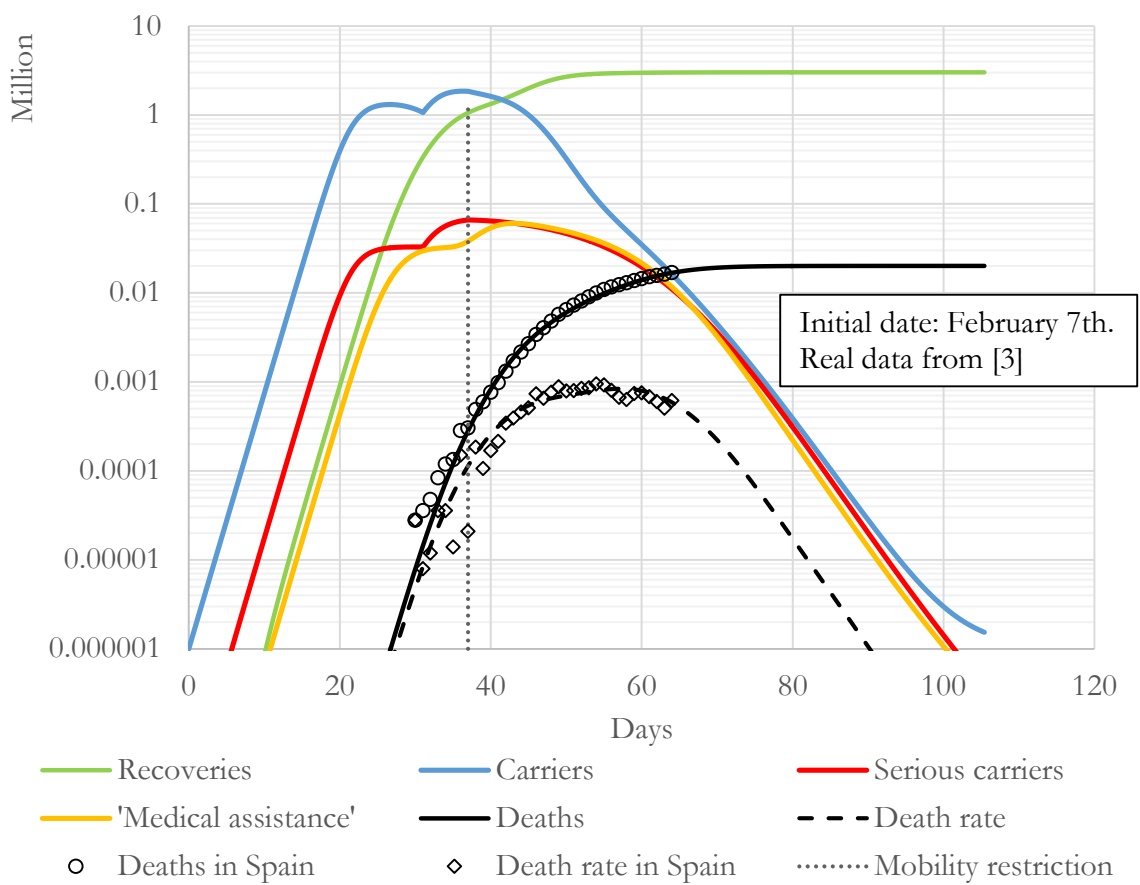

Figure 4. Graphs showing the calculations made with the three initial dates discussed in the text. The parameters used in each graph are collected in Table 1. The three calculations share the following input parameters: $\mu_{IH} = 3.10$ days, $\mu_{OH} = 11.36$ days, $\mu_{I1} = 6.72$ days, $\mu_{I2} = 13.92$ days, $t_D = 0.283$ and $\gamma = 0.165$. Real data comes from reference [3].

## VI. Discussion

Again, it has been shown that the main source of uncertainty is the initial date or, equivalently, the fraction at risk. Plausible values for both parameters have been used in this article but it is important to remark that, according to the presented model and to the available information, they are only plausible values. With these values, the model have shown its ability to theoretically describe the death rate in Spain. The punctual increment of the susceptible population (obtained by fitting) keeps the results close to the actual data: no other actions have been necessary. So far, the proposed differential equations hold and the input parameters are still meaningful. However, further study is still necessary to assess the potential of the new model.